\documentclass[prd, twocolumn, nofootinbib, floatfix, superscriptaddress]{revtex4}
\input epsf
\usepackage{amsmath,amssymb,subfigure}
\usepackage{graphicx}
\usepackage{epsfig}
\usepackage{color}
\usepackage{aas_macros}

\newcommand{\beq}{\begin{equation}}
\newcommand{\eeq}{\end{equation}}
\newcommand{\beqa}{\begin{eqnarray}}
\newcommand{\eeqa}{\end{eqnarray}}
\newcommand{\om}{\Omega_m}

\newcommand{\omde}{\Omega_{\rm{de}}}
 
\newcommand{\lcdm}{\Lambda{\rm CDM}} 
 
\newcommand{\ls}{\mathrel{\raise0.27ex\hbox{$<$}\kern-0.70em \lower0.71ex\hbox{{
$\scriptstyle \sim$}}}}

\def\zph{z_{\rm ph}}
\def\zbias{z_{\rm bias}}
\def\zmed{z_{\rm med}}
\def\neff{n_{\rm eff}}
\def\fsky{f_{\rm sky}}
\def\sz{\sigma_z}
\def\erf{{\rm erf}}
\def\vecn{{\bf{n}}}
\def\gammaadd{\gamma_{\rm add}}
\def\hmpc{$h^{-1}$Mpc}

\def\tot{{\rm{tot}}}
\def \anl{A_{\rm NL}}

\usepackage{hyperref}
\begin{document} 

\title{Weak Lensing Science, Surveys, and Systematics} 
\author{Sudeep Das}
\affiliation{Berkeley Center for Cosmological Physics, University of California, 
Berkeley, CA, USA} 
\author{Roland de Putter} 
\affiliation{IFIC, Universidad de Valencia-CSIC, Valencia, Spain} 
\affiliation{Institut de Ciencies del Cosmos, Barcelona, Spain} 
\author{Eric V.\ Linder} 
\affiliation{Berkeley Center for Cosmological Physics, University of California,
Berkeley, CA, USA}
\affiliation{Institute for the Early Universe, Ewha Womans University, Seoul, Korea} 
\affiliation{Lawrence Berkeley National Laboratory, Berkeley, CA 94720 USA}
\affiliation{Space Sciences Laboratory, University of California, Berkeley, CA  94720, USA}
\author{Reiko Nakajima} 
\affiliation{Berkeley Center for Cosmological Physics, University of California,
Berkeley, CA, USA}
\affiliation{Institute for the Early Universe, Ewha Womans University, Seoul, Korea}
\affiliation{Lawrence Berkeley National Laboratory, Berkeley, CA 94720 USA}
\affiliation{Space Sciences Laboratory, University of California, Berkeley, CA  94720, USA}
\date{\today}

%%%%%%%%%%%%%%%%%%%%%%%%%%%%%%%%%%%%%%%%%%%%%%%%%%%%%%%%%%%%
%%%%%%%%%%%%%%%%%%%%%%%%%%%%%%%%%%%%%%%%%%%%%%%%%%%%%%%%%%%%
\begin{abstract} 
Weak gravitational lensing is one of the key probes of the cosmological 
model, dark energy, and dark matter, providing insight into both the 
cosmic expansion history and large scale structure growth history. 
Taking into account a broad spectrum of physics affecting growth -- 
dynamical dark energy, extended gravity, neutrino masses, and spatial 
curvature -- we analyze the cosmological constraints.  Similarly we 
consider the effects of a range of systematic uncertainties, in shear 
measurement, photometric redshifts, and the nonlinear power spectrum, 
on cosmological parameter extraction.  We also investigate, and provide 
fitting formulas for, the influence of survey parameters such as 
redshift depth, galaxy number densities, and sky area.  Finally, we 
examine the robustness of results for different fiducial cosmologies.  \end{abstract} 

\maketitle

%%%%%%%%%%%%%%%%%%%%%%%%%%%%%%%%%%%%%%%%%%%%%%%%%%%%%%%%%%%%
%%%%%%%%%%%%%%%%%%%%%%%%%%%%%%%%%%%%%%%%%%%%%%%%%%%%%%%%%%%%
\section{Introduction \label{sec:intro}}

Weak gravitational lensing is a key probe of cosmology, in terms of 
both the properties of spacetime as a whole and the distribution of 
matter within the universe \cite{munshi/etal:2008, huterer:2010}.  
It has the characteristic of being 
sensitive to both the cosmic expansion and the growth of structure, 
making it ideal for testing the consistency of these two windows on 
the universe, 
e.g.\ as predicted by general relativity.  Moreover, weak lensing 
has substantial complementarity with other probes, enabling more 
precise constraints through breaking degeneracies. 

Because of this richness, weak lensing not only probes many aspects 
of physics but is also affected by them.  Neglecting some of the 
sources impacting growth of structure could bias the results, 
or at best underestimate the true uncertainties.  In this article we 
present a series of analyses calculating the effects on weak lensing 
science when incorporating simultaneously the major possibilities for 
modification of $\lcdm$ growth.  These include using dynamical dark 
energy, extended gravity, neutrino masses, and spatial curvature.  
This generalizes investigations that only consider one or two of 
these physical ingredients at a time. 

We also examine how conclusions on survey properties alter when the 
fiducial model is different from $\lcdm$, and in the presence of 
systematics in the photometric redshifts, shear measurement, and 
nonlinear power spectrum form.  In Sec.~\ref{sec:base} we present 
our methodology, including the basic equation for the weak lensing 
shear power spectrum and some hidden assumptions in its derivation, 
the set of cosmological parameters, and the framework of our Fisher 
analysis code.  Section~\ref{sec:sys} outlines the survey parameters 
and systematic models, including a parametrized form for the nonlinear 
matter power spectrum.  Cosmological parameter constraints are discussed 
in Sec.~\ref{sec:param}, including effects of an extended 
parameter set, the fiducial cosmology, survey parameters, and 
systematics.  We also present a selection of trade studies and the 
resulting fitting functions, that could guide more effective survey 
design.

%%%%%%%%%%%%%%%%%%%%%%%%%%%%%%%%%%%%%%%%%%%%%%%%%%%%%%%%%%%%
%%%%%%%%%%%%%%%%%%%%%%%%%%%%%%%%%%%%%%%%%%%%%%%%%%%%%%%%%%%%
\section{Foundations and Methodology \label{sec:base}}

\subsection{Weak Lensing Shear} 

We briefly review the key equations for the weak lensing shear power 
spectrum.  For a comprehensive treatment, 
see \cite{bartelmann/schneider:2001, hoekstra/jain:2008}.  The 
shear (equivalently the convergence in the weak lensing limit) is a 
weighted integral of the mass density field. The average convergence
of a light ray bundle from sources 
in a source bin $i$, in the sky direction $\vec\theta$, is
\beq 
\kappa_i(\vec\theta)=\int_0^{\infty} d\eta\,W_i(\eta)\,\delta(\eta, 
\eta\vec\theta)\,, \label{eq:kappa}  
\eeq 
where $\delta=\delta\rho/\rho$ is the fractional mass density 
perturbation and $\eta$ is the conformal time (the upper limit will be 
cut off by the kernel).  
The convergence
weight function 
or kernel is
\beq
W_i(z) = \frac{3}{2}\om H_0^2 (1+z)\, \eta(z) \, \int_z^\infty dz'\,\frac{n_i(z')}{n_i^A} 
\frac{\eta(z,z')}{\eta(z')}.
\eeq
Here the redshift $z$ is another way of measuring the conformal distance 
$\eta$, $\eta(z')$ is the distance to the source, $\eta(z,z')$ the 
distance between lens and source and $\eta(z)$ the distance to the lens.  
The number of sheared objects (in source bin $i$) along 
the line of sight per unit sky area and per unit redshift is given by the source 
distribution $n_i(z)$, with the total number per unit sky area $n_i^A = \int dz\,n_i(z)$.
Finally, the matter density in units of the critical density 
is $\om$, and $H_0$ is the Hubble constant. 

By Fourier transforming to multipole space and forming the two point 
correlation, one obtains the power spectrum $C_\ell$ (omitting source bin subscripts for now): 
\beqa 
\kappa(\vec{\theta})&=&\int \frac{d^2\vec{\ell}}{(2\pi)^2}\, \kappa(\vec{\ell}) \, e^{i \vec{\ell} \cdot \vec{\theta}}\\ 
\langle\kappa(\vec{\ell}) \kappa(\vec{\ell}')\rangle&=&(2\pi)^2\delta^D(\vec\ell+ 
\vec\ell')\, C_\ell\,, 
\eeqa 
in analogy to the mass power spectrum $P_k$, 
\beq 
\langle\delta(\vec{k})\delta(\vec{k}')\rangle=(2\pi)^3\delta^D(\vec k+\vec k')\, P_k\,. 
\eeq 
Note that $\delta^D$ denotes a Dirac delta function and not a density 
perturbation and we employ the flat sky approximation. 

From the above equations one can see that, in the Limber approximation, the shear power spectrum 
between two samples (source redshift bins) $i$ and $j$ is 
\beq
C_{\ell,ij}=\int_0^\infty \frac{d\eta}{\eta^2}\,W_i(z)\,W_j(z)\,P_k(k=\ell/\eta;z) 
\label{eq:cldef}.
\eeq
One immediately sees that weak lensing is sensitive to several distinct 
quantities: cosmic distances in terms of $\eta(z)$ and $\eta(z,z')$ and 
hence the expansion, the present physical mass density $\om H_0^2$, 
the matter power spectrum $P_k(z)$, which involves the expansion history 
$\om(a)$, the growth history, and the linear to nonlinear density mapping, 
and the source distribution $n_i(z)$, which will require translating 
photometric redshifts into true (``spectroscopic'' redshifts), and allows 
the use of redshift tomography or crosscorrelations between distinct 
samples at different stages of growth. 

A number of assumptions are implicit, and often forgotten, in the 
derivation of 
Eq.~(\ref{eq:cldef}).  To obtain the relation between the convergence 
and the matter density perturbation in Eq.~(\ref{eq:kappa}) one solves 
the geodesic equation in general relativity 
specialized to a Friedmann-Robertson-Walker (FRW) spacetime.  A change in the 
gravitational light deflection law, such as through unequal time-time 
and space-space metric potentials, will alter the result, as will a 
modification of the Poisson equation relating the potential to the density 
perturbation.  Properly, the kernel $W$ should be multiplied by the 
$\mathcal{G}$ function generalizing Newton's constant (see 
\cite{daniel/linder:2010}).  See \cite{acquaviva/baccigalupi/perrotta:2004} 
for more details; here we assume that 
both relations take their standard form.  Likewise we assume that only 
matter clusters significantly.  One could also explore the 
effects of taking the light geodesic in a slightly inhomogeneous rather 
than FRW spacetime, i.e.\ one where the light rays received by the observer 
preferentially traverse underdense regions.  This then changes the 
distance factors from the FRW to the Dyer-Roeder form 
\cite{dyer/roeder:1973}.  While this is formally required 
for consistency, practically it is insignificant for weak lensing 
\cite{linder:2008}.  Finally, in the conversion from the homogeneous mass 
density to $\om H_0^2$ one assumes $\rho_m(a)\sim a^{-3}$, that is 
conservation of matter (no interactions with dark energy, for example). 
To summarize, the 
standard formula must be corrected when investigating cosmological 
models involving certain types of gravity, an inhomogeneous universe, 
or matter interactions.  We will keep all the 
standard assumptions.

%%%%%%%%%%%%%%%%%%%%%%%%%%%%
\subsection{Physical Ingredients} 

In weak lensing the cosmological model enters through both the 
expansion history (distances) and the growth history (matter power 
spectrum $P_k$).  The cosmological parameter estimation must therefore 
take into account all parameters that could impact these.  On the expansion 
side this includes not only the matter density but dynamical dark energy 
through its time varying equation of state, and the possibility of spatial 
curvature.  On the growth side in addition to the usual scalar perturbation 
power index one must consider again the effects of dynamical dark energy and spatial 
curvature, plus neutrino mass and the possibility of growth modifications 
through extensions to general relativity. 

Thus the vanilla $\lcdm$ cosmological parameter set must be enlarged to 
account for these physical effects.  We include the dynamical dark energy 
equation of state $w(a)$ through the two parameters $w_0$ and $w_a$, where 
$w(a)=w_0+w_a(1-a)$ where $a=1/(1+z)$, the spatial curvature density $\Omega_k$, 
the sum of neutrino masses $\sum m_\nu$, and modified growth through the gravitational 
growth index $\gamma$, especially suitable for gravitational modifications 
that are scale independent on the scales relevant for weak lensing. 
Dark energy suppresses growth due to the increased Hubble expansion rate 
and smooth spatial distribution (we do include dark energy perturbations but 
this contributes little to the matter power spectrum).  Neutrino mass 
suppresses growth through free streaming.  Spatial curvature acts as an 
unclustered component and so effectively dilutes the matter clustering. 
Gravitational modifications can enhance or suppress growth.  Cosmological 
parameter estimation when including only one of these effects can lead to 
incorrect conclusions, if others exist as well, due to the similarity of 
their physical impacts.

%%%%%%%%%%%%%%%%%%%%%%%%%%% 
\subsection{Weak Lensing Code}
\label{sec: wlcode} 
\begin{table*}[t]
\caption{\label{tab:params}Parameter Set and Fiducial Values}
\begin{minipage}{\textwidth}
\begin{center}
\begin{tabular}{|lc|cc|} 
\hline
\hline 
\multicolumn{2} {|c|}{Cosmological} & \multicolumn{2}{c|}{Systematic} \\
\hline
Parameter & Value &  Parameter & Value\\
\hline
\multicolumn{2} {|c|}{$\lcdm$} & \multicolumn{2}{c|}{Photometric} \\
\hline
$\Omega_b h^2$ & 0.02258  & $z_{\rm{pz}}^{k}$& \{0.0, 0.3,   0.6,   0.9,   1.2,   1.5,   1.8,   2.1,   2.4,   2.7,  3.0\}\footnote{The photo-z systematics are defined through linear interpolation between these 11 redshift nodes.} \\
$\Omega_c h^2$ & 0.1109 & scatter, $\sigma_z^{k}$ & $0.03~(1+z_{\rm{pz}}^{k})$\\
$\omde$ & 0.734 & bias, $z_{\rm{bias}}^{k}$   &  $0.0~(1+z_{\rm{pz}}^{k})$\\
\cline{3-4}
$n_s$ &0.96 &\multicolumn{2}{c|}{ Additive Shear\footnote{Defined through Eq.~(\ref{eq:shearadd}).}  (5 tomographic bins)} \\
\cline{3-4}
$\sigma_8$ &0.8 & $\alpha'$ & 1.0 \\
\cline{1-2}
\multicolumn{2} {|c|}{Beyond-$\lcdm$} & $\rho$ & 0.5 \\
\cline{1-2}
$\sum m_\nu$ &0.15  eV\footnote{Three degenerate species assumed.}&$\ell_*$& 1000\\
$\Omega_k$  &0.0 & $b^i$ & \{$10^{-5}$, $10^{-5}$, $10^{-5}$, $10^{-5}$, $10^{-5}$\}\\
\cline{3-4}
$\gamma$ & 0.55 &\multicolumn{2}{c|}{ Multiplicative Shear\footnote{Defined through Eq.~(\ref{eq:shearmult}).} (5 tomographic bins)}\\
\cline{3-4}
$w_0$ &-1.0 & $f_i$  & \{0.0, 0.0, 0.0, 0.0, 0.0\}\\
$w_a$  &0.0\footnote{When $w_0$ and $w_a$ are varied,  their fiducial  values are slightly offset so as to  prevent $w(a)$ crossing $-1$.}&&\\
\hline
\hline
\end{tabular}
\end{center}
\end{minipage}
\end{table*}

The weak lensing cosmology code developed by the authors is possibly unique 
in that it includes all four of the cosmological influences on growth, 
together with inclusion of observational systematics.  This paper presents 
examples of analysis using this Berkeley weak lensing code to study the 
impact of these additional contributions.  The code uses a modified version 
of CAMB \cite{lewis/challinor/lasenby:2000} to include the additional parameters in calculating the linear 
matter power spectrum, Halofit generalized by the inclusion of the growth 
index $\gamma$ to form the nonlinear matter power spectrum, and then 
integrates over the geometric and source distribution kernel to compute the 
weak lensing shear power spectrum.  Cosmological 
parameter estimation is carried out through Fisher matrix analysis, with 
the error covariance matrix including systematics in photometric redshifts 
of the source galaxies being lensed and in the shear measurements. 
We also consider systematics in the form of the nonlinear mass power spectrum 
in Sec.~\ref{sec:nonlin}. 

The Berkeley weak lensing code will be made publicly available in the near future. 

The cosmological parameter set for the Fisher matrix analysis includes 
the vanilla $\Lambda$CDM parameters, namely, the density parameter for baryons 
$\Omega_b h^2$,  for dark matter $\Omega_c h^2$,  and  for dark energy $\omde$, 
as well as the scalar spectral index $n_s$ and the amplitude of matter fluctuations 
at redshift $z=0$ on 8~\hmpc\ scales, $\sigma_8$. Beyond-$\Lambda$CDM  parameters 
explored in this paper include the growth index $\gamma$, the sum of neutrino masses 
$\sum m_\nu$, spatial curvature parameter $\Omega_k$, and the dark energy 
equation-of-state parameters $w_0$ and $w_a$.  Although not discussed in this paper, 
the code also includes the following parameters: the number of relativistic species 
$N_{\rm eff}$, reionization optical depth $\tau$, and the running of the spectral index 
$\alpha$, in anticipation of synergistic studies with CMB and CMB lensing data sets.  

Apart from the cosmological parameters, a set of parameters describing systematic 
effects can be varied. These  are further discussed in Sec.~\ref{sec:sys}.  
Table~\ref{tab:params}  displays the cosmological and systematic parameters and 
their fiducial values. 

Information on parameters $\{p_\mu\}$ can be obtained,  
under the assumptions of Gaussianity of the observables, 
from the Fisher matrix 
\begin{equation}
F_{\mu\nu} = \sum_{\ell = \ell_{\rm{min}}}^{\ell_{\rm{max}}} \frac{\partial X_{\ell,a}}{\partial p_\mu}  \left[ ({\rm{Cov}}_\ell)^{-1}_{ab}\right]\frac{\partial X_{\ell,b}}{\partial p_\nu}, 
\end{equation}
where to simplify notation we define 
\beq
X_{\ell, a = i(i-1)/2+ j}\equiv C_{\ell, ij} \hspace{2mm} (i\ge j )\,.
\eeq 
The covariance matrix is given by
\beq
({\rm{Cov}}_\ell)_{ab}  = \frac 1{(2\ell+1) f_{\rm{sky}}}\left[ C_{\ell,ik}^{\tot}C_{\ell,jl}^{\tot}+C_{\ell,il}^{\tot}C_{\ell,jk}^{\tot}\right], 
\eeq
with $a \equiv i(i-1)/2+ j$ and $b \equiv  k(k-1)/2+ l$. The total spectra $C_\ell^{\tot}$ 
are defined as the sum of the cosmological signal and noise: 
\beq
C_{\ell,ij}^{\tot} = C_{\ell,ij} + N_{\ell,ij} \label{eq:noise} 
\eeq 
where  the noise spectra $N_{\ell}$ encode the effects of shape noise and other 
systematics, as discussed in Sec.~\ref{sec:sys}. For the purpose of  this study, we 
have set $\ell_{\rm{min}}= 2$ and $\ell_{\rm{max}}=3000$, the latter being the maximum 
multipole up to which the Gaussian approximation is sufficiently valid.  Here $f_{\rm sky}$ 
is the fraction of sky covered by the experiment; we chose $f_{\rm sky}= 0.097$ 
corresponding to 4000 square degrees.

%%%%%%%%%%%%%%%%%%%%%%%%%%%%%%%%%%%%%%%%%%%%%%%%%%%%%%%%%%%%
%%%%%%%%%%%%%%%%%%%%%%%%%%%%%%%%%%%%%%%%%%%%%%%%%%%%%%%%%%%
\section{Systematics \label{sec:sys}} 

The number of galaxies whose shears are measured in a large survey can be many 
millions so the statistics are copious and systematic uncertainties play a  
correspondingly larger role.  We do not present a comprehensive analysis (see, 
for example, \cite{bernstein:2009}) -- many 
of the effects depend on observational details -- but include the two standard 
sources of uncertainty coming from the data and a further one from theory. 

%%%%%%%%%%
\subsection{Photometric Redshift Systematics}
\label{sec:photoz}

Obtaining accurate, spectroscopic redshifts for the many millions of measured galaxies 
is impractical, so weak lensing surveys rely on photometric redshifts.  These will 
be imperfectly calibrated by a spectroscopic subsample, and have a residual scatter 
and bias with respect to the true redshifts.  Since this alters the kernel in the 
shear power spectrum, these systematics will propagate into the cosmological parameter 
estimation. 

We follow the photometric redshift model as described in \cite{ma/hu/huterer:2006}.
For an observed photometric redshift $\zph$, the probability distribution $p(\zph|z)$
of the true redshift is modeled as a Gaussian distribution
\begin{equation}
p(\zph|z) = \frac{1}{\sqrt{2\pi}\sz} \exp\left[-\frac{(z-\zph-\zbias)^2}{2\sz^2}\right]
\end{equation}
where $\zbias(z)$ and $\sz(z)$ are the bias and scatter in $p(\zph|z)$, respectively,
and where we allow them to be an arbitrary function of redshift $z$.

We then assume an overall true galaxy redshift distribution $n(z)=d^2N/dz\,d\Omega$ of 
\begin{equation}
\label{eq: numdens}
n(z) \propto z^\alpha \exp\left[-(z/z_0)^\beta\right]
\end{equation}
where we have implemented $\alpha=2$ and $\beta=1$; 
$z_0$ is a characteristic redshift which is related to the median redshift by
$z_0 = z_{\rm med} / 2.674$ in that case, and 
the normalization is fixed by the total number of galaxies per steradian
\begin{equation}
n^A = \int_0^\infty dz\, n(z).
\end{equation}
The true distribution for objects binned in photometric redshifts, with lower to upper limits
of $\zph^{(i)}$ to $\zph^{(i+1)}$, is then
\begin{eqnarray}
n_i(z) &=& \int_{\zph^{(i)}}^{\zph^{(i+1)}} d\zph\, n(z) p(\zph|z) \\
&=&\frac{1}{2} n(z) [\erf(x_{i+1}) - \erf(x_{i})]
\end{eqnarray}
where $x_i$ is defined as
\begin{eqnarray}
x_i \equiv \left( \zph^{(i)} - z + \zbias \right) /\sqrt{2} \sz.
\end{eqnarray}
The total number of galaxies per bin per steradian is then
\begin{equation}
n^A_i = \int_0^\infty dz\, n_i(z).
\end{equation}

We identify $\zbias$ and $\sz$ as the nuisance parameters for the Fisher matrix analysis,
where they are taken to be $z$ dependent.  These values are defined at {\tt npzbin} intervals,
where intermediate values are linearly interpolated.  Hence there are $2\times${\tt npzbin} 
nuisance parameters associated with photometric redshift-related systematics.
It is the uncertainties in these parameters, more than the values of the parameters per se, 
that impact the cosmological parameter estimation.

%%%%%%%%%%
\subsection{Shear Systematics}
\label{sec:shearsys}

Measurement of the shear from galaxy images is a complicated process, since galaxies 
have (unknown) intrinsic shape ellipticities and imperfect resolution and optical 
distortions due to the telescope, 
detectors, and atmosphere contribute as well.  The galaxy shapes can be treated statistically 
through a shape noise contribution $\sigma_\gamma^2/n^A_i$ in Eq.~(\ref{eq:noise}) but the other 
uncertainties give residual systematics. 

We follow the formalism in \cite{huterer/etal:2006} for additive and multiplicative 
shear systematics in the weak lensing signal extracted from the galaxy images. 
The multiplicative systematic is typically generated from the incomplete removal of the
finite-size effects of the point spread function (PSF), but can also be generated
from non-ideal weighting of galaxy shapes to estimate the net shear.  
Its effect can be encapsulated in
the factor $f$ 
\begin{equation}
\hat{\gamma}(z, \vecn) = \gamma(z, \vecn)\,[1+f(z, \vecn)]
\end{equation}
where $\hat\gamma(z, \vecn)$ and $\gamma(z, \vecn)$ are the estimated and true shear at
some (true) redshift $z$ and direction $\vecn$.  Notice that $f$ is itself is a function of
$z$ and $\vecn$; it is also time dependent (due to the PSF dependence on these quantities).
Within the tomographic bin $i$, we average the multiplicative factor $f$ over $\vecn$ and $z$
such that we are left with the bias $f_i\equiv\langle f(z_i,\vecn)\rangle$.
The observed shear correlation function between the $i$th and $j$th tomographic bin is then
\begin{eqnarray}
 \langle \hat\gamma(z_{\rm i},\vecn)\hat\gamma(z_{\rm j},\vecn + d\vecn)\rangle 
\label{eq:shearmult}\hspace{0.5\columnwidth} \\
=\langle \gamma(z_{\rm i},\vecn)\, \gamma(z_{\rm j},\vecn + d\vecn)\rangle (1+f_i) (1+f_j) 
\nonumber \\ 
\simeq \langle \gamma(z_{\rm i},\vecn)\, \gamma(z_{\rm j},\vecn + d\vecn)\rangle (1+f_i+f_j)\,. \hspace{3mm}
\nonumber 
\end{eqnarray}

The additive systematic is typically generated from the anisotropy of the PSF, 
and is characterized by
$\gammaadd$, 
\begin{equation}
\hat{\gamma}(z, \vecn) = \gamma(z, \vecn) + \gammaadd(z, \vecn) \,.
\end{equation}
Since the PSF anisotropy is uncorrelated with the true shear $\gamma$, we can assume
$\langle \gamma(\vecn)\, \gammaadd(\vecn+d\vecn) \rangle = 0$.  The non-vanishing term is then
$\langle \gammaadd(\vecn)\, \gammaadd(\vecn+d\vecn) \rangle$, with Legendre transform 
$P^\kappa_{\rm add}(\ell)$.  The additive shear systematic then affects the weak lensing 
angular power spectra as
\begin{equation}
\hat{C}_{\ell,ij} = C_{\ell,ij} + N_{\ell,{\rm{add}}} \,. 
\end{equation}
It is the power spectrum of the residual after characterization of the additive shear 
that impacts the observations.  Hence the additive shear error is given by 
\begin{equation}
\gammaadd(z_i,\vecn) = b_i r(\vecn) \,, 
\end{equation}
where $r(\vecn)$ is a direction dependent random variable, and $b_i$ is the characteristic
additive shear residual amplitude.  The additive shear error power spectrum is assumed to 
take the form of a power law \cite{huterer/etal:2006},  
\begin{equation}
\label{eq:shearadd}
N_{\ell,\rm add} = \rho b_i b_j \left(\frac{\ell}{\ell_*}\right)^{\alpha'} \,, 
\end{equation}
where the coefficient $\rho$ is set to unity for $i=j$, and fixed to some fiducial value for 
$i\neq j$.  
The choice of $\ell_*$ is arbitrary and is degenerate with the parameters $b_i$;
we choose $\ell_* = 1000$ since the weak lensing signal is generally sensitive to 
this region.  We take $\alpha'=1$; 
\cite{huterer/etal:2006} find their results are insensitive to this power law index. 

We include $f_i$, $b_i$, and $\alpha'$ as nuisance parameters for the Fisher matrix analysis. 
The parameters $f_i$ and $b_i$ are defined for each tomographic bin. 
Hence there are $2\times${\tt ntom}$+1$ nuisance parameters associated with shear systematics,
where {\tt ntom} is the number of tomographic bins.  

%%%%%%%%%%%%%%%% 
\subsection{Nonlinear Power Spectrum \label{sec:nonlin}} 

Weak lensing accesses information from both the linear and nonlinear 
matter density regimes.  
The nonlinear matter density power spectrum is a key element in the 
shear power spectrum, but the contribution of the nonlinearities is  
imperfectly known.  Frequently the nonlinear power spectrum is 
generated from the linear power spectrum through the Halofit 
prescription \cite{smith/etal:2003}, but even for $\Lambda$CDM this is 
accurate at only the $\sim10\%$ level.  A detailed and valuable  
investigation of the effects of systematic uncertainties in the 
nonlinear power spectrum on cosmological parameter estimation from 
weak lensing was carried out by \cite{huterer/takada:2005}. 

They considered the density power spectrum as piecewise constant in 
bins of wavenumber $k$ and found that each bin needed to be constrained 
to $\sim1\%$ level in order not to degrade cosmology estimation from 
next generation weak lensing surveys.  The worst bias caused by the 
systematic uncertainty comes from step features in the power spectrum, 
i.e.\ sharp changes with $k$.  Many physical influences on 
the power spectrum, however, would be expected to be smooth with wavenumber. 

Here, we consider one specific case of a smooth multiplicative function 
representing the systematic uncertainty.  The form adopted is roughly motivated 
by modified gravity effects on the matter spectrum as found from 
both analytic arguments and numerical simulations \cite{oyaizu/lima/hu:2008}.  
Although inspired by $f(R)$ gravity models with a chameleon mechanism 
restoring general relativity on small scales, the form should simply be 
taken for itself, representing a deviation from the $\Lambda$CDM 
prediction of Halofit on small scales, then a constant offset on even 
smaller scales out to the $k_{\rm max}$ limit included in the weak 
lensing calculations.  (There might be a restoration to zero offset 
on even smaller scales past our limit.) 

The example parametrization is 
\beq
\label{eq: ansatz 1}
P(k) = P_{\rm HF}(k) \, \left(1 + \frac{A_{\rm NL} \left(k/k_0\right)^2}{1 + \left(k/k_0\right)^2} \right),
\eeq
where $k_0 = 1 \, h\,$Mpc$^{-1}$, $A_{\rm NL}$ is a free parameter and $P_{\rm HF}$ is
the non linear power spectrum as given by Halofit \cite{smith/etal:2003}. 

The power spectrum systematic enters through assuming the wrong value of $A_{\rm NL}$, 
e.g.\ using Halofit only, that has $A_{\rm NL}=0$, 
when there is really a modification.  We first consider the 
systematic as a bias $\delta\anl=\anl{}_{\rm assumed}-\anl{}_{\rm true}$.  This propagates 
into a bias in the cosmological parameters through 
the Fisher bias formula.  In general, if we have $N + 1$ parameters 
$\{p_i\}$ and we misestimate the parameter $p_{N + 1}$ by $dp_{N + 1}$, 
the resulting bias in $p_i$ is 
\beq
\label{eq: bias}
\frac{d p_i}{d p_{N + 1}} = - \sum_{j = 1}^N (F^{(N)})^{-1}{}_{i j}\, 
(F^{(N + 1)})_{j, N + 1}\,. 
\eeq
In our case $\delta p_{N + 1} = \delta \anl$. 

For our fiducial weak lensing survey, combined with Planck CMB 
information, the multiplicative leverage of 
a power spectrum systematic is a factor of 18 (6) for $w_a$ ($w_0$). 
That is, a 10\% misestimation of $\anl$ biases $w_a$ by $1.8\sigma$. 

Alternatively, if one is sure of the form (\ref{eq: ansatz 1}) for the power spectrum, one 
can try to fit for $\anl$.  In this case, the other parameters will not be biased, but 
their errors might degrade.  In fact, we find that 
for the weak lensing plus Planck CMB combination, the degradation is 
less than 5\% in the dark energy parameters and the power spectrum 
uncertainty can be determined to $\sigma(\anl) = 0.016$.  
This highlights the crucial role of understanding the form of the $k$ 
dependence of the nonlinear matter power spectrum.

%%%%%%%%%%%%%%%%%%%%%%%%%%%%%%%%%%%%%%%%%%%%%%%%%%%%%%%%%%%%
%%%%%%%%%%%%%%%%%%%%%%%%%%%%%%%%%%%%%%%%%%%%%%%%%%%%%%%%%%%%
\section{Cosmology Constraints and Survey Design \label{sec:param}} 

\subsection{Effect of Systematics \label{sec:cossys}} 

In this section, we investigate the effect of the various systematics discussed in the previous
section on the cosmology constraints. We consider a fiducial survey with a sky coverage of
$4000$ deg$^2$ and an effective projected galaxy number density 
$\neff = 55\, {\rm arcmin}^{-2}$.  
We take the intrinsic galaxy shear to be $\sigma_\gamma = 0.27$.  
We assume a redshift distribution given by Eq. (\ref{eq: numdens}) with
$z_{\rm med} = 1$
and subdivide the galaxy sample into five redshift bins, given by
$z = 0.6 - 1.0 - 1.5 - 2.1 - 3.9$. It was shown in \cite{ma/hu/huterer:2006} that five tomographic
bins is sufficient to capture the redshift information.

For the photometric redshift errors (see Sec.~\ref{sec:photoz}),
we assume a fiducial redshift scatter \cite{ma/hu/huterer:2006}
\beq
\sigma_z(z) = 0.03 \, (1 + z)
\eeq
and our fiducial bias is zero. To characterize the uncertainty in the scatter and bias,
we parametrize $\sigma_z(z)$ and $z_{\rm bias}(z)$ by their values in
{\tt npzbin}$=11$ bins, include these $22$ parameters in the Fisher matrix,
apply a prior of 0.01 on each, and finally marginalize over these parameters. 

For the additive shear bias (see Sec.~\ref{sec:shearsys}), we assume
fiducial values $b_i = 10^{-5}$ for each tomographic bin. Since the true additive shear
is unknown (if it were known, we would not consider it a systematic), we again add the
$b_i$ as parameters in the Fisher matrix, with priors of $10^{-5}$. 
Finally, we consider the multiplicative shear systematic (see Sec.~\ref{sec:shearsys}), 
with fiducial parameters $f_i = 0$ and priors of $0.01$.

\begin{figure}[!t]
\includegraphics[width=\columnwidth]{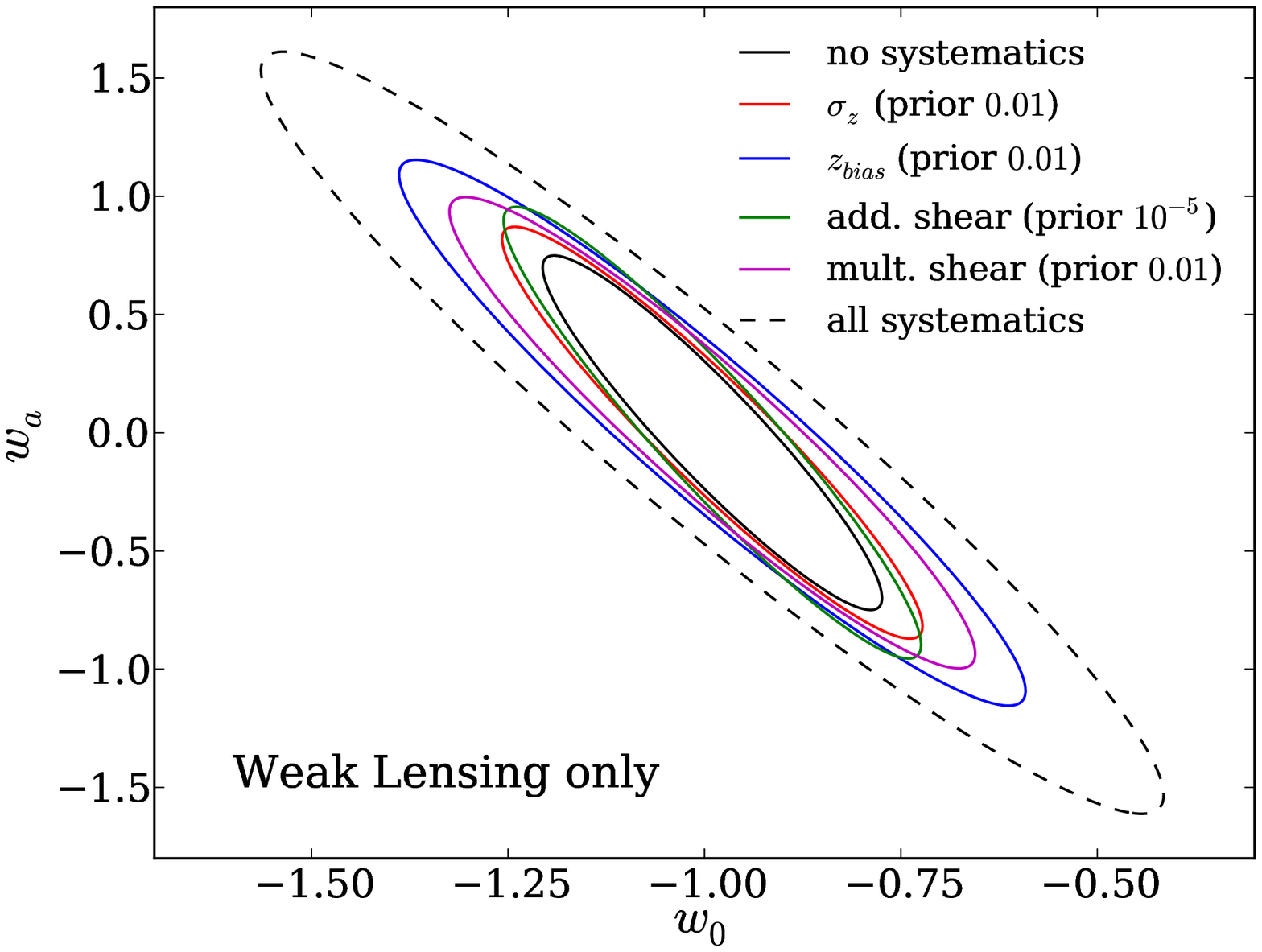}
\includegraphics[width=\columnwidth]{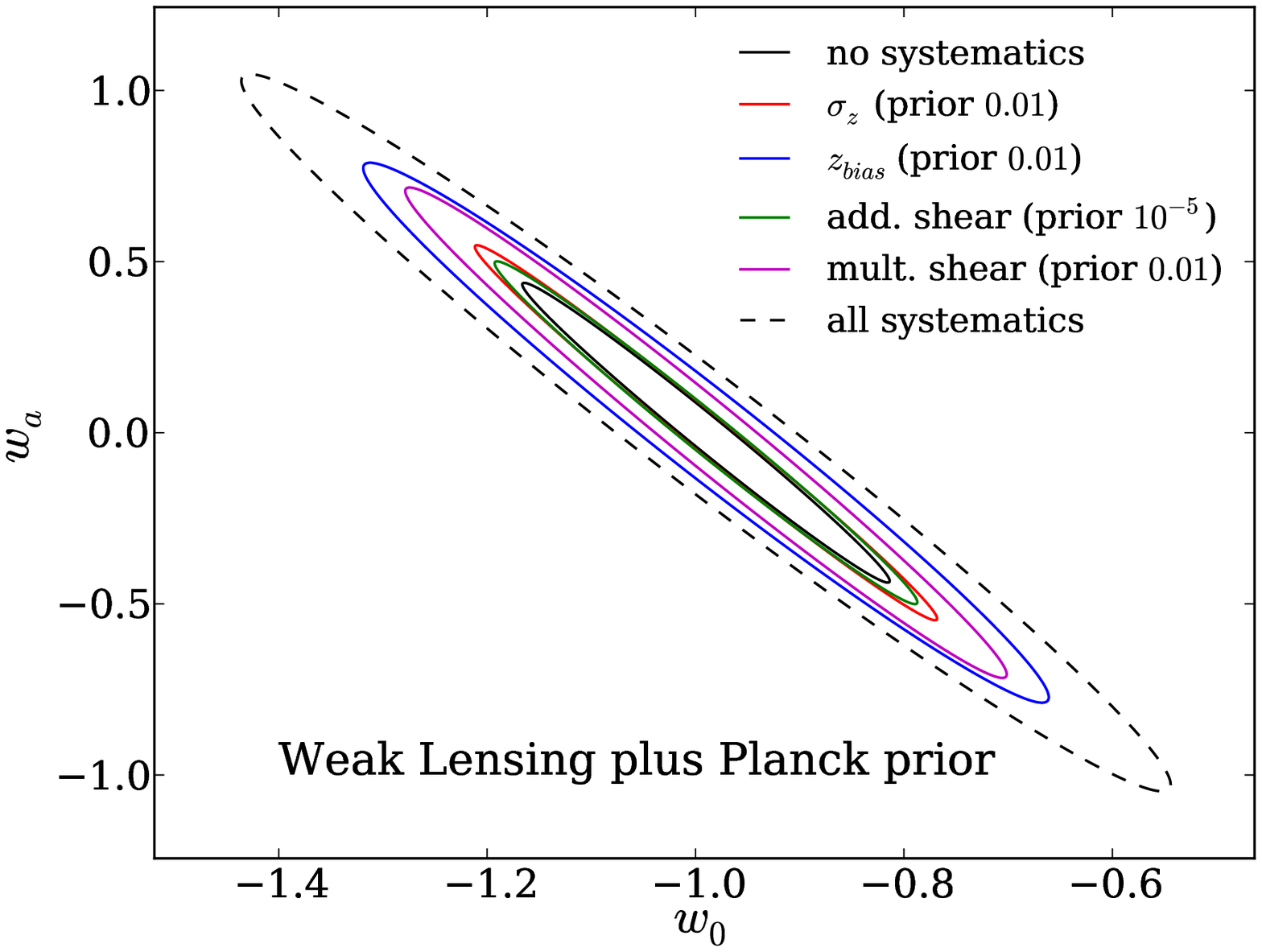}
\caption{
Dependence of the dark energy equation of state constraints 
($68 \%$ CL contours) on the following systematics:
uncertainty in the photometric redshift scatter and bias, multiplicative shear
and additive shear. We show contours where each effect is taken into account individually,
a contour without systematics, and one with all four systematics included.
For the priors chosen here, the multiplicative shear and the photometric
redshift bias are the dominant systematics. Taking into account all of them,
errors are increased by about a factor of 2, and the contour area (``figure of merit'') 
by about a factor of 5.  In the bottom panel only, we also include our code-generated forecasted Planck CMB Fisher matrix.
}
\label{fig: allsys}
\end{figure}

In Fig.~\ref{fig: allsys}, we show the effects of the systematics on constraints for 
the dark energy parameters $w_0$ and $w_a$, marginalized over all other parameters. 
We also include a forecasted Planck CMB prior in the bottom panel only, showing how 
that information breaks certain degeneracies (mostly between parameters other than 
$w_0$ and $w_a$, but this results in narrowing the $w_0$-$w_a$ contours).  

In either case, the systematics have noticeable effects, even for a 4000 square degree 
weak lensing survey, on the dark energy constraints.  We show the effects of each 
systematic individually, as well as all of them included simultaneously.  The strongest 
impact comes from the multiplicative shear and the photometric redshift bias uncertainty, 
at least for the prior levels we have considered as reasonable expectations (although 
one should be able to improve on $z_{\rm bias}$ with a good spectroscopic calibration set 
for the photometric redshifts).   
In general, we expect the systematics contribution
is largest from the multiplicative shear error and redshift bias since these components
most directly affect the calibration of the lensing signal, and hence cosmological parameters.
The results agree qualitatively with 
\cite{ma/hu/huterer:2006, huterer/etal:2006}, so we now proceed to study the extension to a 
larger array of cosmological physics affecting large scale structure growth.

%%%%%%%%%%%%%%%%%%%%%%%%%%%%%%%%%%%%%%%%%%%%%%%%%%%%%%%%%%%%%%%%%%%%%%%%%%%
\subsection{Effect of Cosmological Physics \label{sec:cosphys}}

\begin{figure}
\includegraphics[width=\columnwidth]{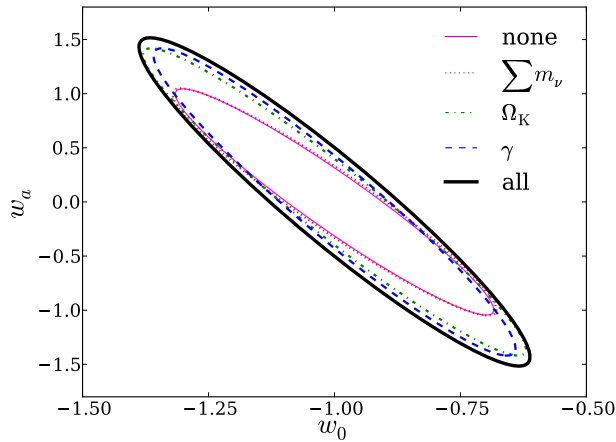}
\caption{Dependence of the dark energy constraints on beyond-$\lcdm$ parameters: the sum 
of neutrino masses $\sum m_\nu$, spatial curvature $\Omega_k$ and the growth index $\gamma$. 
The 68\% contours are shown for the case where none of these is varied, when each parameter 
is marginalized over individually, and when all of them are marginalized over. \label{fig:cosmoSys} }
\end{figure}

As mentioned in the introduction, several distinct types of physics affect cosmic growth.  
In this section we investigate the effects of beyond-$\Lambda$CDM parameters on dark 
energy constraints.  Besides the dynamical dark energy equation of state itself, we 
consider three parameters describing these influences --- the spatial 
curvature density parameter $\Omega_k$, the sum of the masses of neutrinos $\sum m_\nu$, 
and the gravitational growth index $\gamma$.  The effects of varying each of these 
individually and all at the same time are shown in Fig.~\ref{fig:cosmoSys}.  
Allowing for variation of all three parameters together leads to a worsening of the 
dark energy figure of merit (FOM) by a factor of $\sim 2$.  

Either relaxing flatness (allowing for $\Omega_k\ne0$) or that growth must follow 
expansion (allowing for $\gamma\ne\gamma_{\rm GR}$) have large impacts.  For curvature, 
recall that this enters into both the distances and growth part of the weak lensing 
signal; generally combining accurate distance information, from a supernova or baryon 
acoustic oscillation survey, would ameliorate its impact on the cosmology constraints. 
Allowing for growth to deviate from the general relativity prediction (with matter being 
the dominant clustering component), weakens the dark energy constraints, but guards 
against substantial bias if such deviations do exist \cite{huterer/linder:2007}.  For example, a 
departure $\Delta\gamma$ can give rise to a bias $\Delta w_a\approx 8\Delta\gamma$. 
Although this extra freedom generically increases the $w_0$-$w_a$ contour area by 
factor of $\sim 2$, the equation of state constraints end up being nearly independent 
of the actual value of $\gamma$ (see Fig.~2 of \cite{linder:2011}). 

Remarkably, varying the sum of neutrino masses only decreases the figure of merit 
only by $\sim 7\%$. This is mainly due to the fact that neutrino free streaming suppresses 
growth of structure in a uniquely scale dependent way, with growth on smaller scales (higher $k$) being more 
suppressed than on larger ones, while the effects of dark energy,  growth index or curvature have no particular 
scale dependence. As a consequence, 
the correlation coefficients between the uncertainties in neutrino mass and in dark 
energy equation of state parameters are small (see, e.g., Table 2 of \cite{stril/cahn/linder:2010}). 

\begin{figure}
\includegraphics[width=\columnwidth]{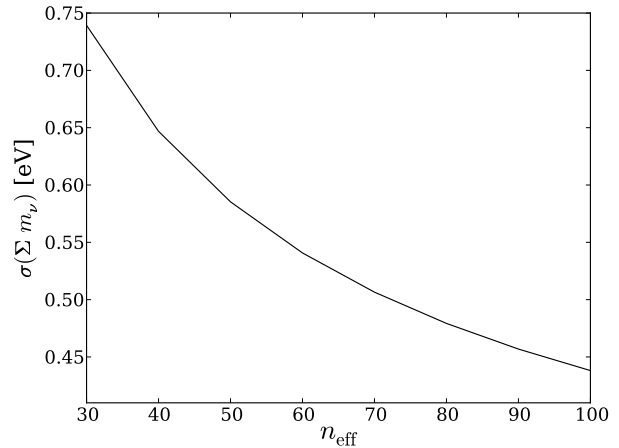}
\caption{Dependence of the constraint on the sum 
of neutrino masses $\sum m_\nu$ on the source density $n_{\rm eff}$ for the $4000$ sq. deg. 
survey considered here.  \label{fig:mnuVsNbar}}
\end{figure}

Since this scale dependent suppression of growth makes	
weak lensing power spectra sensitive to the neutrino
mass sum, we here investigate the constraints, from
weak lensing alone, on the sum of neutrino masses.
Figure~\ref{fig:mnuVsNbar} shows the dependence of this
constraint, marginalized over all other parameters in
the $w_0$-$w_a$ cosmology, on the source density of
sheared galaxies for fixed survey area (4000 sq.~deg.).	
Increased source
density lowers the uncertainty on the shear power spectra,
especially on smaller scales where the effects of neutrino
streaming are more pronounced. 
We find changing the sky area scales the constraint by
$\fsky^{1/2}$.

%%%%%%%%%%%%%%%%%%%%%%%%%%%%%%%%%%%%%%%%%%%%%%%%%%%%%%%%%%%%%%%%%%%%%%%%% 
\subsection{Redshift Range Contribution \label{sec:zdepth}} 

One of the interesting issues to explore is the redshift range over 
which weak lensing measurements contribute significantly to cosmological 
parameter determination.  This feeds directly into the survey design trade 
off of a wide area but shallower survey vs a narrower but deeper survey. 
We investigate this in terms of both the redshift dependence of the 
kernel support for the weak lensing shear power spectrum (i.e.\ the 
lens redshift), and in terms of the maximum source galaxy redshift. 

Since the shear power spectrum is an integral over the line of sight 
(Eq.~\ref{eq:cldef}), we can analyze the redshift dependence of the 
integrand, or kernel of the projected shear power.  Recall that $z(\eta)$ 
is the lens redshift.  In particular, we expect that for small redshifts 
the contribution goes to zero since $\eta$ gets small and for a fixed 
multipole $\ell$ the scales probed get small and the mass power leading 
to shear decreases.  (Alternately, if one considers constant 
fractional distance intervals $d\eta/\eta$, then the path length $\eta$ 
itself gets small.)  For high lens redshifts, the lens-source distance 
$\eta_{ls}\equiv\eta(z,z')$ gets small and again the integrand vanishes.  Thus the 
kernel support peaks at intermediate lens redshifts, although the peak 
does shift to higher redshifts as the source distribution, i.e.\ $\zmed$, 
does.  (Note that the rule of thumb that lensing is strongest at a 
conformal distance midway between the source and observer is more directly 
relevant to the lensing {\it cross-section\/} than to the shear power 
contribution.) 

\begin{figure}[!t]
\includegraphics[width=\columnwidth]{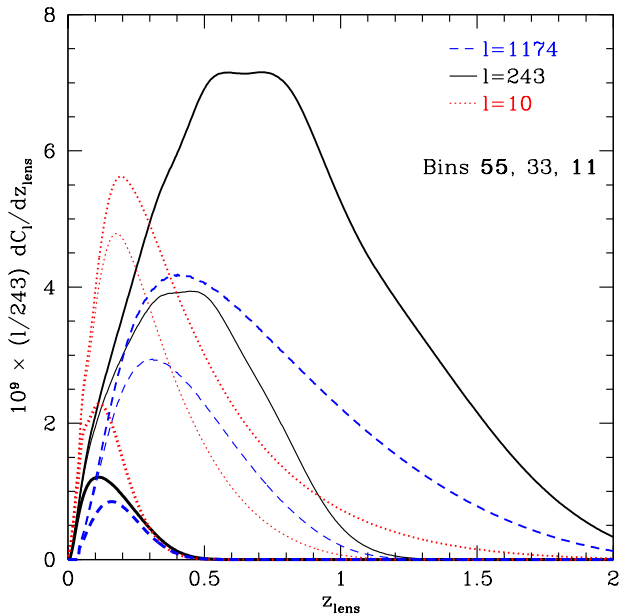}
\caption{
The contributions of each redshift interval to the shear power 
spectrum, i.e.\ the kernel support, are plotted vs lens redshift 
to show over what range the power arises.  We plot $\ell$ times the shear 
autopower in source redshift bins 1, 3, and 5 to show 
the different multipoles with roughly the same amplitude.  The 
interplay of the peak in the mass power $P_k$ and the geometric factor 
$\eta_{ls}$ means that high multipoles eventually get stronger 
contributions from smaller redshifts, like low multipoles do, while 
intermediate multipoles probe higher redshifts. 
}
\label{fig:kernel}
\end{figure}

Figure~\ref{fig:kernel} illustrates the redshift dependence of the 
shear power integrand for several values of multipole $\ell$ and 
source redshift bin.   We show small, medium, and large $\ell$ and 
the autopower in bins 1 ($0<z<0.6$), 3 ($1<z<1.5$), and 5 ($2.1<z<3.5$). 
Note that the peak for $\langle z_s\rangle\approx0.3$ (1.25, 2.8) is at 
$z_{\rm lens}\lesssim0.2$ (0.4, 0.7).  There is a long tail though to 
higher lens redshifts.  Turned around, one could say that to probe 
gravitational lensing in the universe out to $z\approx0.5$ requires 
sources out to $z\gtrsim2$.

\begin{figure}[!t]
\includegraphics[width=\columnwidth]{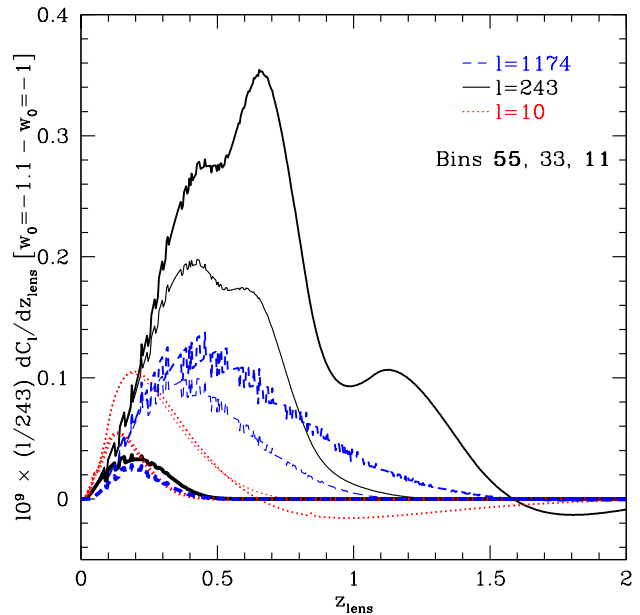}\\ 
\includegraphics[width=\columnwidth]{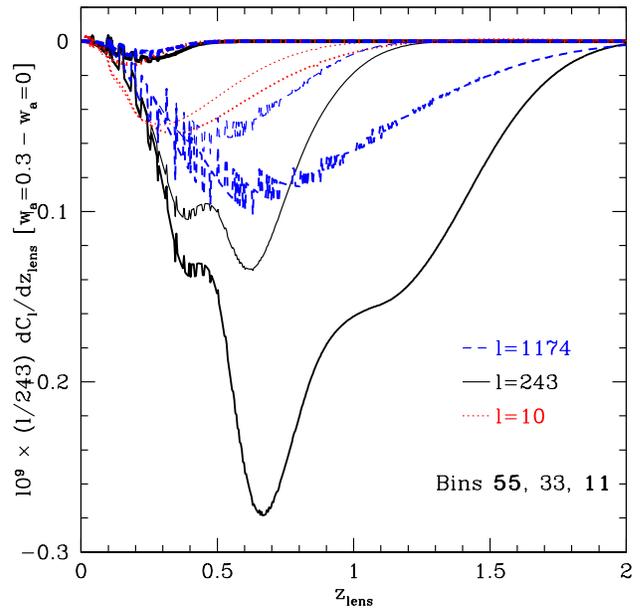}
\caption{
The contributions of each redshift interval to the cosmological parameter 
discrimination ability of the shear power spectrum, with respect to $w_0$ 
(top panel) and $w_a$ (bottom panel), are plotted vs lens redshift to show 
from which redshifts the leverage on the parameters 
arises.  The jitter in the curves is numerical noise in the derivatives. 
We multiply the finite difference between two cosmologies of the shear 
autopower in source redshift bins 
1, 3, and 5 by $\ell$ simply to plot the different multipoles with roughly 
the same amplitude.  The redshift dependence of the leverage is similar to, 
or weighted to slightly higher redshifts, than that of the shear power itself. 
}
\label{fig:dCdw}
\end{figure}

For cosmological constraints, what is of greater importance than where 
the shear power peaks is where lies the innate information about the 
cosmological parameters.  That is, does information on different 
cosmologies exhibit a similar redshift weighting as the overall 
shear power integrand, or does higher redshift data, for example, 
show greater cosmological leverage?  Figure~\ref{fig:dCdw} examines this 
issue with respect to the dark energy equation of state parameters 
$w_0$ and $w_a$.

Considering the finite difference between the kernel supports $dC_\ell/dz$ 
for cosmologies with two different parameter values (all other 
parameters held fixed), e.g.\ effectively $dC_\ell/(dz\,dw_0)$, the 
contributions are seen to be essentially comparable regarding the peak redshift 
to the shear power itself.  For $w_a$, there is a slight shift in 
leverage to 
higher peak redshifts, e.g.\ the peak for $\ell=243$, bin 3 is $z\approx0.6$ 
rather than 0.45.  
The persistence of dark energy to higher redshifts through  
a positive $w_a$ increases the cosmology signal at high redshift, while 
a more negative $w_0$ diminishes the high redshift dark energy density 
and decreases the signal at high redshift.  
In both cases, however, the rise from small $z$ is pushed to larger 
$z$ than in the power kernel.  This makes sense since low redshift 
cosmology is insensitive to the value of $w_0$ and even more so $w_a$, 
and so the leverage is delayed.  

We examine the impact of the survey redshift depth on the dark energy 
equation of state figure of merit in Sec.~\ref{sec:cosfid}, where we 
also investigate the interaction with fiducial cosmology.

%%%%%%%%%%%%%%%%%%%%%%%%%%%%%%%%%%
\subsection{Source Density and Sky Area \label{sec:cossky}} 

Another important aspect of survey design concerns the number density 
of source galaxies.  The large number density of galaxies useful for 
shear measurements that become available with a small point spread 
function is one of the major advantages of space-based observations \cite{kasliwal/etal:2008}. 
The density $\neff$ involves several factors, including the PSF 
characterization and the redshift depth (since the number density is 
projected along the line of sight, plus the galaxy size distribution 
varies with redshift). 

\begin{figure}[!htbp]
\includegraphics[width=\columnwidth]{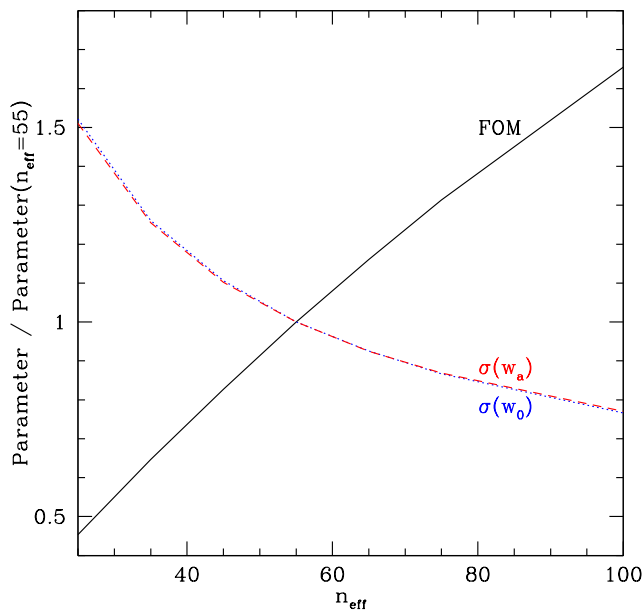}
\caption{
The dark energy figure of merit improves significantly with source density 
$\neff$ of measurably sheared galaxies. Halving the fiducial source density 
of 55 measured galaxies per square arcminute halves the FOM. 
}
\label{fig:fomn}
\end{figure}

Figure~\ref{fig:fomn} shows that greater source density has a strong 
effect on figure of merit, with almost a linear dependence in the 
range of $\neff$ of interest (the improvement does flatten out at 
higher values, as the shape noise term becomes negligible).  This study 
treats $\neff$ as an isolated parameter, and does not account for the 
accompanying gain in PSF characterization likely to come from a higher 
density of sources and star calibrations; it also holds the survey depth 
$\zmed$ fixed.  The fitting function over the range plotted is given by 
\beq 
\frac{{\rm FOM}(\neff)}{{\rm FOM}(\neff=55)}=0.016\neff+0.12 \,. 
\eeq

The total number of lensed sources also scales with the sky area, or 
fraction $\fsky$.  More sources implies more information and so the figure 
of merit increases with sky area.  The systematics associated with weak 
lensing, in photometric redshifts and in shear measurements, are not 
taken to provide a systematic floor traditionally, i.e.\ the shear 
power spectrum uncertainty has a factor $1/\fsky$ outside all the error 
contributions and so continues to drop with ever greater area.  
Whether this is realistic or not is an open question.  
However, priors that are placed on marginalized parameters accounting for 
these systematics do break the linear scaling. 

We examine the dependence of the dark energy parameter estimation and 
figure of merit on the sky area for the particular case of a prior on 
the photometric redshift bias of $10^{-2}$.  This should be a fairly 
generic representative of the behavior among the systematics.  
For relatively small survey 
area, the statistical leverage of the survey data itself provides the 
limit on cosmological parameter estimation, from fitting for all the 
systematics, and the redshift bias prior is not so important.  With 
large sky area, however, the other contributions are well enough 
determined that they become of the same order as the prior on $z_{\rm bias}$ 
and so the prior plays an important role.  See, for example, Fig.~7 of 
\cite{ma/hu/huterer:2006}.  The result is that the cosmological parameter estimation 
(FOM) slows down in its improvement with $\fsky$, although it does not 
reach a ceiling. 

\begin{figure}[!htbp]
\includegraphics[width=\columnwidth]{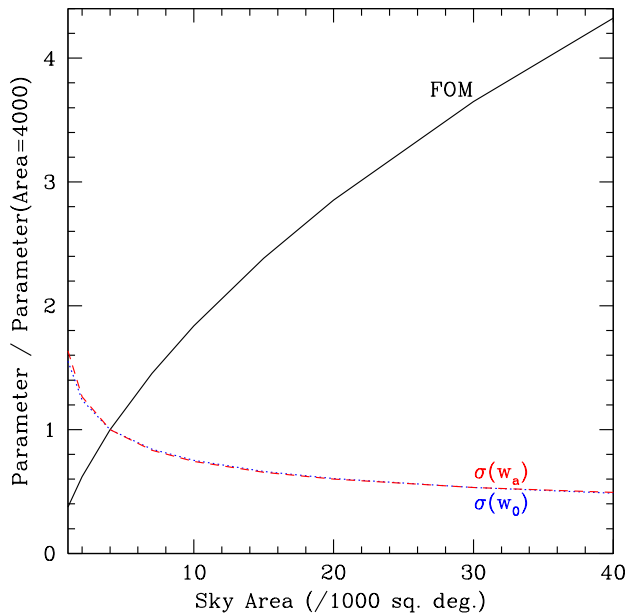}
\caption{
In the presence of priors on systematics (or a systematic floor), the 
cosmological parameter estimation does not improve linearly with sky 
area, but somewhat more slowly.  
}
\label{fig:fomf}
\end{figure}

Figure~\ref{fig:fomf} illustrates the change in scaling.  Note that 
even for a survey area of 1000 deg$^2$ the improvement goes more slowly 
than linearly with $\fsky$.  In the cosmological parameter determination 
there is a noticeable bend in the scaling between 4000 and 10000 deg$^2$, 
after which the slope is shallower.  A fitting function is 
\beq 
\frac{{\rm FOM(area)}}{{\rm FOM(4000\, deg}^2)}= 
\left(\frac{A}{4000}\right)^{1-0.316(A/4000)^{0.0615}} \,, 
\eeq 
where sky area $A$ is in units of square degrees.

%%%%%%%%%%%%%%%%%%%%%%%%%%%%%%%%%%
\subsection{Effect of Fiducial Model \label{sec:cosfid}}

One aspect we do not have control over in the survey design is the 
true cosmological model.  In the projections for how well a survey will 
constrain the cosmological parameters, the true values of the parameters 
can have an impact.  Here we examine not only the extent of the influence on 
the parameter estimation but whether one would be led to optimize survey 
design differently depending on what one believed the true cosmology was. 

\begin{figure}[!htbp]
\includegraphics[width=\columnwidth]{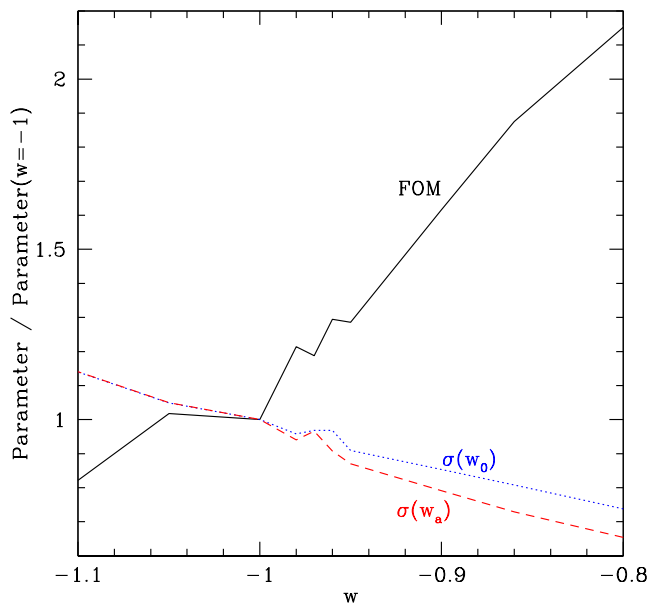}
\caption{
Cosmological parameter estimation improves as the effective dark energy 
equation of state becomes less negative.  The area figure of merit (FOM) 
improves by a factor 2 for a shift of fiducial cosmology from $w=-1$ to 
$w=-0.84$.  (Jitter in the curves near $w=-1$ is an artifact of adjusting 
numerical step sizes to prevent crossing of $w=-1$.) 
}
\label{fig:fomw}
\end{figure}

In general, one expects that as the dark energy persists to higher 
redshift, e.g.\ if its equation of state is closer to zero, its effects 
should be more visible in the shear power spectrum and it should be 
better constrained.  Figure~\ref{fig:fomw} shows that this is indeed 
the case.  For a shift from $w=-1$ to less negative values, 
the area figure of merit in the dark energy equation of state $w_0$-$w_a$ 
plane increases by 
\beq 
\frac{{\rm FOM}(w)}{{\rm FOM}(w=-1)}-1\approx 6\,(1+w)\,. 
\eeq 
That is, a shift $\Delta w=0.1$ boosts the FOM by a factor 1.6.

Note that here $w$ denotes the effective constant dark energy equation 
of state, defined for the dynamical equation of state case 
using the rule of thumb that the effective value $w=w_0+w_a/3$. 
We have verified that this holds true, e.g.\ the results for $w_0=-0.97$, 
$w_a=0.21$ agree with $w=-0.9$ to better than 4\%, and likewise for 
$w_0=-0.8$, $w_a=0.18$ and $w=-0.86$.  For fiducial models with $w$ more 
negative than $-1$, the change in FOM has less strong dependence since 
there is little contribution from the higher redshift bins regardless of 
the exact value of $w<-1$.  The scatter in the figure curves near $w=-1$ 
is a numerical artifact from interaction of the small step size (so as 
not to cross $w=-1$) with the dark energy perturbation evolution. 

\begin{figure}[!htbp]
\includegraphics[width=\columnwidth]{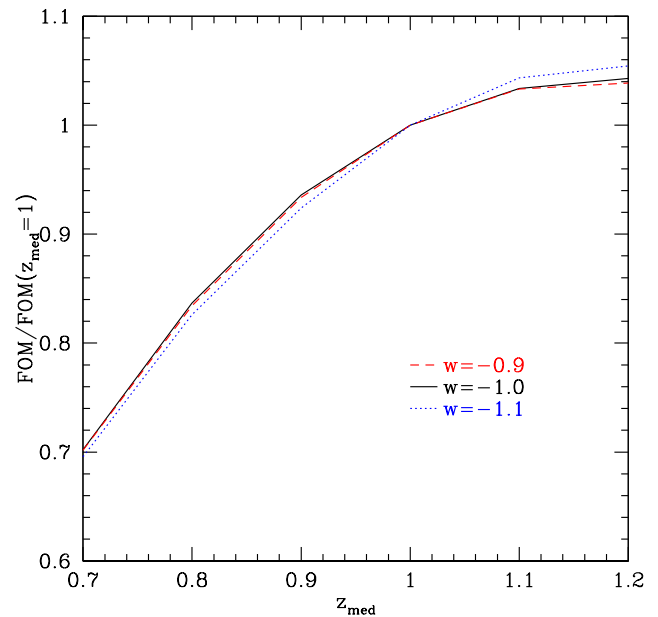}
\caption{
The dark energy figure of merit improves significantly with survey depth, 
up to $\zmed\approx1$, and then nearly saturates.  This example of survey 
design optimization is robust against assumption of the fiducial cosmology, 
with the dependence on $\zmed$ being similar for $w=-0.9$ to $-1.1$. 
}
\label{fig:fomz}
\end{figure}

These results show that weak lensing surveys may have more probative 
power than expected from taking a $\Lambda$CDM fiducial cosmology. 
However, we have no control over what the true cosmology is, so what we 
do need to check is how sensitive conclusions about survey design are 
to the fiducial model.  We would not necessarily want to choose survey 
requirements that are near optimal for one cosmology but inefficient for 
another.  Figure~\ref{fig:fomz} demonstrates that the survey requirements, 
with regards to the survey depth $\zmed$ at least, are robust with respect 
to fiducial cosmology.

The first important result is that the survey depth $\zmed$ is a crucial 
factor in the cosmological power of the survey.  A survey whose median 
source depth is $\zmed=1$ has a 40\% improvement in FOM over one with 
$\zmed=0.7$, all else held fixed.  However, going deeper than 
$\zmed\approx1.0$ does not lead to any continued substantial gain.  Thus, 
$\zmed\approx1.0$ is near optimal for survey design.  The second important 
point is that this optimum is not sensitive to the (uncontrollable) true 
cosmology.  Models with effective dark energy equations of state between 
$w=-0.9$ and $-1.1$ all follow this rule, with the FOM as a function of 
$\zmed$ (normalized to the $\zmed=1$ case) differing by less than 3\% 
between them. An excellent fit for the FOM as a function of survey depth 
over the range plotted is 
\beq 
\frac{{\rm FOM}(\zmed)}{{\rm FOM}(\zmed=1)}=1.04- 
0.04\left(\frac{1.2-\zmed}{0.2}\right)^{2.34} \,. 
\eeq

%%%%%%%%%%%%%%%%%%%%%%%%%%%%%%%%%%%%%%%%%%%%%%%%%%%%%%%%%%%%%%%%%%%
\section{Conclusions \label{sec:concl}} 

Weak gravitational lensing can be a powerful probe into the 
cosmological model, testing both the expansion history and 
growth of large scale structure.  This sensitivity also requires 
that a broad array of physics be included when calculating the 
weak lensing shear power spectrum and its constraints on 
cosmological parameters.  We include the effects of dynamical dark 
energy, spatial curvature, neutrino masses, and gravitational 
modifications into a weak lensing Fisher code, together with 
systematics from photometric redshift measurements, shear 
measurements, and nonlinearities in the mass power spectrum. 

This flexibility allows us to carry out studies analyzing the 
effects of each element of an expanded parameter space, of 
systematic uncertainties, and of survey characteristics on the 
ability for weak lensing to probe cosmology.  We show that allowing 
growth of structure some independence from the expansion history, 
through the gravitational growth index $\gamma$, i.e.\ relaxing the 
dictum of general relativity, enlarges the dark energy equation 
of state $w_0$-$w_a$ contour area by a factor 2, but could lead 
the detection of new aspects of gravity.  We find that including 
a systematic uncertainty in the nonlinear regime behavior of the 
matter power spectrum has little effect -- {\it if\/} the $k$ 
dependence of the modification is known; ignoring the uncertainty 
(or getting the dependence wrong) can lead to substantial bias in 
cosmology. 

Systematics control in photometric redshifts and shear measurement 
are, as is well known, critical.  The cumulative effect of these 
uncertainties is especially powerful.  The systematics feed strongly 
into the survey design, for example an improvement in shear measurement 
allowing a higher effective number of galaxies $\neff$ yields a 
substantial improvement in the dark energy figure of merit.  

We have presented a number of illustrative survey characteristic 
analyses, giving accurate fitting formulas for the effect on the 
dark energy figure of merit with source density, survey depth, 
and sky area.  An interesting point is that the fiducial cosmology 
can have a substantial impact on the figure of merit, but fortunately 
the behavior with survey characteristics (at least as far as depth) 
is nearly independent of fiducial model.  Thus survey design can 
focus on systematics, within the control of experimental, 
simulation, and theory efforts.  Given these efforts, weak lensing 
promises a cornucopia of knowledge about cosmology, gravity, and 
fundamental physics.

\acknowledgments

We thank Dragan Huterer for useful discussions.  
This work  has been supported in part by the 
Director, Office of Science, Office of High Energy Physics, of the 
U.S.\ Department of Energy under Contract No.\ DE-AC02-05CH11231, and 
also by the World Class University grant R32-2008-000-10130-0 
through the National Research Foundation, Ministry 
of Education, Science and Technology of Korea (EL,RN).  RN acknowledges 
partial support by NASA LTSA Carlo 23235-24523-44.

%%%%%%%%%%%%%%%%%%%%%%%%%%%%%%%%%%%%%%%%%%%%%%%%%%%%%%%%%%%%
%%%%%%%%%%%%%%%%%%%%%%%%%%%%%%%%%%%%%%%%%%%%%%%%%%%%%%%%%%%%
%\begin{thebibliography}{99}

%\end{thebibliography} 

\end{document}